# ¿Apropiación privada de renta de recursos naturales? El caso del cobre en Chile*

## Private Appropriation of Natural Resource Rent? The case of Cooper in Chile

*Benjamín Leiva Crispi*\*\*

ABSTRACT

Unexpected increases of natural resource prices can generate rents, value that should be recovered by the State to minimize inefficiencies, avoid arbitrary discrimination between citizens and keep a sustainable trajectory. As a case study about private appropriation of natural resource rent, this work explores the case of copper in Chile since 1990, empirically analyzing if the 12 main private mining companies have recovered *in present value* more than their investment during their life cycle. The results of this exercise, applicable to other natural resources, indicate that some actually have, capturing about US$40 billion up to 2012. Elaborating an adequate institutional framework for future deposits remain important challenges for Chile to plentifully take advantage of its mining potential, as well as for any country with an abundant resource base to better enjoy its natural wealth. For that purpose, a concession known as Least Present Value Revenue (LPVR) is proposed.

Resumen

Los aumentos de precio de un recurso natural pueden generar renta, cuyo valor debiese ser recuperado por el Estado para minimizar ineficiencias en la economía, evitar discriminaciones arbitrarias entre ciudadanos y mantener una trayectoria sostenible. Como tema de estudio sobre apropiación privada de renta de recursos naturales, este artículo explora el caso del cobre en Chile desde 1990, analizando empíricamente si las principales mineras privadas han recuperado *en valor presente* más que su inversión. Los resultados del ejercicio, aplicable a otros recursos naturales, indican que algunas mineras sí lo han hecho, captando unos 40 000 millones de dólares hasta 2012. Elaborar una institucionalidad idónea para futuros yacimientos es un importante desafío para que Chile aproveche su potencial minero, así como para que cualquier país con una abundante base de recursos goce plenamente de su riqueza natural. Para tal efecto, se propone la concesión conocida como Mínimo Valor Presente de Ingresos (mvpi).

*Palabras clave:* cobre, recursos naturales, renta, análisis intertemporal, Mínimo Valor Presente de Ingresos, concesiones, Chile. *Clasificación JEL:* D43, L72, Q39.

Introducción

Una particularidad de las industrias que extraen recursos naturales es que son proclives a generar renta: ganancias extraordinarias sobre el retorno necesario para inducir a que empresas inviertan en ellas (Barma *et al.,* 2012). Con el objetivo de captar la renta para el Estado, muchos países poseen normativas especiales, constituidas por impuestos específicos, concesiones u otros mecanismos, esfuerzo justificado por motivos de eficiencia, equidad y sostenibilidad.

Un importante aspecto de la discusión sobre la renta de recursos naturales es su existencia. En términos generales, si la tasa de retorno de una explotación resulta superior a una estimación generosa del costo de oportunidad del capital *en su sector respectivo,* puede concluirse con seguridad que ha existido renta. No obstante, un análisis adecuado de renta requiere de un enfoque de ciclo de vida de cada explotación, conforme a la posibilidad de considerar debidamente el costo de oportunidad del capital.

Las concesiones son un mecanismo ampliamente usado para gobernar la explotación de recursos naturales, consistente en la definición de un conjunto de derechos y deberes entre quien extrae el recurso (en adelante el agente) y el Estado. Dentro de los derechos figura el acceso exclusivo al recurso en cuestión y la protección de la inversión materializada; y en los deberes, el cumplimiento de estándares ambientales específicos y el pago de impuestos adicionales a los generales del país. La justificación de esto último es la transferencia de la renta del recurso del agente al Estado. Con base en lo anterior, las concesiones de cobre en Chile desde 1990 pueden ser usadas como un caso de estudio sobre la generación de renta de recursos naturales en general, al arrojar lecciones relevantes para el manejo de la renta de metales, hidrocarburos, peces, entre otros.

Basado en un marco teórico de concesiones, este trabajo analiza si las principales mineras privadas cupríferas en Chile han recuperado más que su inversión en valor presente, lo que constituye una definición elemental de renta. Los resultados indican que algunas sí lo han hecho. En la medida en que este análisis prueba ser correcto, Chile ha perdido en eficiencia y equidad, por lo que se en riesgo su sostenibilidad. En consecuencia, la institucionalidad minera que gobierna los yacimientos en Chile resulta inadecuada y requiere una reforma, al tiempo que se convierte en una valiosa fuente de aprendizaje para otros países similares.

Como recomendación de política para futuras concesiones mineras en particular y de recursos naturales en general, se plantea una forma de concesión alternativa ya usada en Chile para concesiones viales, llamada Mínimo Valor Presente de Ingresos (MVPI). Con ciertos ajustes, esta forma podría ser utilizada en cualquier sector que explote recursos naturales y cumpla con ciertas características bien definidas. Por tanto, los razonamientos y propuestas realizados son relevantes para toda economía con una importante base de recursos naturales por desarrollar, como el petróleo de Brasil y México y el cobre de Perú y Chile. La tendencia creciente del volumen de renta revela la importancia de este tema (Banco Mundial, 2010).

La existencia y captación de renta son aspectos cruciales de la discusión en torno al manejo de recursos naturales, pero no son los únicos. La transformación de renta en riqueza es un desafío mayor que involucra primero generarla, luego captarla y finalmente invertirla (Barma *et al.,* 2012). Este trabajo se concentra en la captación de renta, sin ánimo de desconocer la importancia de los otros dos aspectos de un manejo exitoso de la riqueza natural de un país.

El trabajo consta de cinco secciones. La primera realiza una revisión de la bibliografía sobre la renta de recursos naturales, junto a sus alcances en eficiencia, equidad y sostenibilidad. La segunda incorpora una breve presen-tación de la importancia del cobre en la economía chilena para contextualizar al lector y darle significado a las implicaciones del trabajo. La tercera sección presenta un marco teórico para establecer las características de una concesión óptima de recursos naturales en términos de plazos y asignación de renta. La cuarta aplica este marco a las principales empresas privadas cupríferas de Chile para dilucidar si ha existido apropiación privada de renta. Finalmente, en las conclusiones se presentan los resultados y recomendaciones de política.

## I. Renta de recursos naturales

La distinción entre utilidad económica y contable resulta fundamental para comenzar toda discusión sobre manejo óptimo de recursos naturales. El rasgo elemental de la utilidad económica o renta económica (en adelante renta) es que representa la utilidad que queda después de pagar todos los costos, incluyendo el costo del capital —su costo de oportunidad—. En contraste, el concepto de utilidad contable sólo considera los costos explícitos de una actividad cualquiera y por tanto no sustrae el costo del capital invertido.

En el área de recursos naturales, la renta representa el valor económico de dichos recursos, sean cobre, oro, petróleo o carbón, el cual sirve para administrar su escasez (Figueroa, 1999). En efecto, es la escasez de algún factor productivo lo que genera renta en general; y la escasez de un recurso natural (sea de su calidad, volumen limitado, naturaleza agotable, entre otras características) lo que genera renta de recursos naturales en particular (Figueroa, 1999; Harman y Guj, 2006; Otto *et al.*, 2006).

Un aspecto relevante es que no existe renta en mercados perfectamente competitivos. En este contexto, la actividad económica genera utilidad contable igual al costo de oportunidad del capital y, por tanto, el excedente de explotación representa el pago a dicho factor. Toda renta que pudiese generarse se disipa rápidamente por la entrada de nuevos competidores, que disminuyen el precio del producto o aumentan el precio de algún factor productivo limitante.

Por su parte, la renta de recursos naturales no es de la misma naturaleza que aquella de sectores innovadores como los de biotecnología, nanotecnología, farmacéutica, entre otros. En los primeros, la renta deriva de la escasez de recursos naturales y por ello es básicamente *maná del cielo;* en los segundos, la renta es en realidad una cuasi renta, un incentivo necesario para la innovación, el cual está siendo disipado constantemente por innovaciones competidoras (Schumpeter, 1942; Carrol, 2006). No obstante, resulta clave reconocer que para la fase de *exploración* de algunos recursos naturales, tales como metales e hidrocarburos, debe existir un cierto monto de cuasi renta para compensar su alto riesgo, la que puede ser identificada adecuadamente. Sin considerar de manera apropiada esta sutileza, la captación de renta puede llegar a atentar contra la extracción futura del recurso y por tanto contra la generación de renta, al desincentivar futuras inversiones en exploración.

Los Estados suelen tratar sus industrias de recursos naturales de una manera distinta al resto de los sectores de la economía (Daniel *et al.,* 2010). Aunque dicho trato especial puede tener diversos orígenes (Podobnick, 2006), uno de ellos deriva del reconocimiento de que los sectores de recursos naturales no se caracterizan por poseer condiciones de competencia perfecta y en consecuencia sus retornos podrían superar significativamente el retorno requerido para el capital. En este caso, el objetivo del trato particular es precisamente captar la renta que pueda generarse (Gillis, 1982), lo que es importante por motivos de eficiencia, equidad y sostenibilidad.

La captación de renta es relevante en términos de eficiencia, pues permite una recaudación por parte del Estado que reduce las distorsiones en la economía (Boadway y Flatters, 1993; Eggert, 1999). En particular, el hecho de que la recaudación de la renta sea óptima deriva de su naturaleza excedentaria, lo que en definitiva implica que los impuestos que ejecutan dicha recaudación son neutros —no cambian la decisión de agente económico alguno— (Garnaut y Clunies Ross, 1983; Otto, 1995; Figueroa, 1999).

También por motivos de eficiencia, la concesión de recursos es óptima. Como la renta constituye un valor excedentario de la producción, impide que el mercado funcione como un mecanismo discriminador entre firmas eficientes e ineficientes. La imposibilidad de generar competencia *en* la cancha requiere generar competencia *por* la cancha, lo que se traduce en una subasta por el derecho de concesión cuyo valor teóricamente equivaldría a la renta que ésta produjera (Chadwick, 1859; Demsetz, 1968).

En cuanto a eficiencia, el fracaso de captar la renta da espacio para el rentismo, actividad que gasta recursos, pero no contribuye en lo absoluto a la creación de riqueza en la sociedad (Karl, 1997). Por su parte, la captación de renta es importante por motivos de equidad, pues la propiedad sobre los recursos naturales suele ser pública, y cuando no lo es, hay argumentos fundados por los cuales debería serlo (Boadway y Flatters, 1993; Eggert, 1999). Dado esto, al Estado le correspondería una compensación adecuada por su uso, de la misma forma que cualquier propietario cobra por el uso y goce de un activo que le pertenece (Eggert, 1999). Una compensación adecuada corresponde al valor de la renta que produce la explotación del recurso, que, como ya se ha mencionado, representa el valor económico del recurso minero propiamente dicho (Figueroa, 1999). En caso contrario, la privatización de la renta además de ser un hecho económicamente arbitrario, ocasiona problemas de equidad y potenciales conflictos sociales al privilegiar a algunos miembros de la sociedad con ingresos y riqueza que no contribuyeron a crear (Collier, 2007).

Por último, la captura de renta es relevante por motivos de sostenibilidad en cuanto los recursos sean no renovables, por ejemplo minerales e hidrocarburos. Como Hotelling (1931) y Hartwick (1977) demostraron, la manera de lograr una senda de consumo no decreciente en el tiempo cuando se usa un recurso no renovable, es reinvertir toda la renta de tal recurso en otra forma de capital que permita reemplazar el flujo de ingresos una vez que el recurso no renovable se haya agotado. En caso de generarse renta y no reinvertirla, el proceso de extracción mermaría las perspectivas futuras de bienestar del país en cuestión.

A continuación se presenta la importancia del cobre en Chile para darle contexto al lector sobre el caso de estudio utilizado en este trabajo.

## II. La importancia del cobre en Chile

Chile es un país abundante en recursos naturales; destaca especialmente por sus reservas y extracción de cobre. Como muestra la gráfica 1, el cobre ha sido central para el crecimiento económico de Chile durante las últimas décadas. Entre 1970 y 2011 concentró alrededor de 30% de la inversión extranjera directa (IED); casi duplicó al subsector que le sigue (CIE, 2011). Entre 2003 y 2010 representó en promedio más de 15% del PIB; en ocasiones superó 20%. Y respecto a exportaciones, el cobre representó en promedio 49% del total en el periodo 2002-2011 (Cochilco, 2011).

La importancia actual del cobre ha sido consecuencia de un incremento reciente del producto cuprífero nacional. Entre 1991 y 2011, la producción creció 2.9 veces, de 1.8 a 5.2 millones de toneladas métricas de cobre fino (TMCF) (Cochilco, 2011). Como se observa en la gráfica 2, este incremento se debe a grandes inversiones del sector privado, principalmente por empresas extranjeras.

GRÁFICA 1. *Contribución media del cobre*

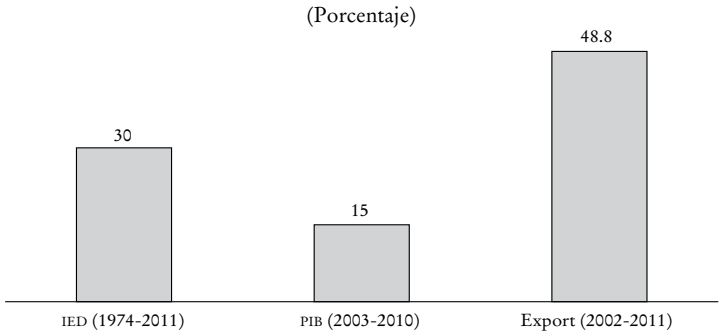

(Porcentaje)

FUENTE: elaboración propia con base en Cochilco (2011) y CIE (2011).

GRÁFICA 2. *Extracción de cobre por minera*

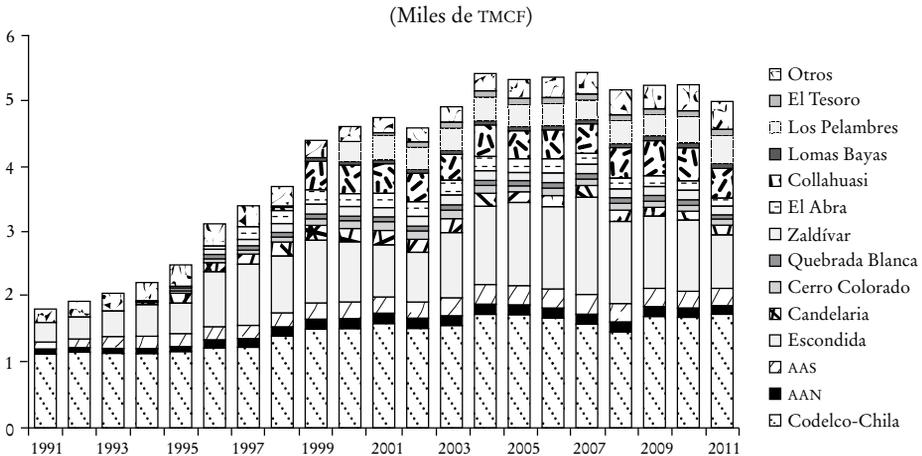

(Miles de TMCF)

FUENTE: elaboración propia con base en Cochilco (2011).

Actualmente, el sector privado representa dos tercios de la extracción total; las operaciones de mayor magnitud son Escondida, Collahuasi y Los Pelambres. Ante esto, cabe preguntarse si la institucionalidad minera que gobierna a estos yacimientos ha sido capaz de captar la renta del cobre para el Estado chileno. Este trabajo intenta responder esta pregunta.

### III. MARCO TEÓRICO

La existencia de renta no puede medirse adecuadamente en términos corrientes si se analizan para años seleccionados los retornos sobre activo o patrimonio. Como Leturia y Merino (2004) señalan, altos retornos en ciertos años podrían estar sólo recuperando pérdidas de periodos anteriores o una inversión realizada hace mucho tiempo. Por este motivo resulta clave un análisis intertemporal de cada yacimiento.

## 1. *El modelo*

Al seguir una versión simplificada y modificada del modelo presentado por Engel y Fischer (2008) para desarrollar un yacimiento (que puede ser de cobre, oro, petróleo, carbón, entre otros recursos) se requiere una Inversión Inicial $I_{t=0}$, que se asume igual entre un gran número de firmas que no se pueden coludir.[1] Dicha inversión rinde una utilidad contable (después de impuestos) que puede ser descrita por una función de densidad $f(U_\tau)$, en que $U_\tau$ es el valor presente *ex post* de la Utilidad Contable, y $f(U_\tau)$ la probabilidad de ocurrencia *ex ante* de cada valor de $U_\tau$ (*i. e.*, la magnitud del riesgo de demanda, operacional y otro tipo de riesgos). Por último, se asume que el recurso extraído se vende en un mercado mundial sobre el cual no hay incidencia posible.

La concesión se caracteriza por un esquema de pagos $V_x(U_\tau)$ que representa la utilidad contable (después de impuestos) en valor presente que recibe el inversionista por aportar su capital. Se asume que el esquema de pagos no incluye ningún tipo de transferencia por parte del Estado al agente, por lo que $0 \leq V_x(U_\tau) \leq U_\tau$, y que el Estado es quien posee los derechos residuales sobre la extracción del recurso (es decir, sobre la renta). Por tanto, para todo valor de $U_\tau$, el Estado recibe la utilidad contable en valor presente equivalente a $U_\tau - V_x(U_\tau)$, lo que corresponde a la renta del recurso, $-R_{x,\tau}-$.

Se asume que para todos los estados posibles del mundo el proyecto es rentable ($U_\tau > I_{t=0}$), que no existe riesgo de expropiación, que $f$ no depende de acción alguna de la firma concesionaria (*i. e.*, no existe riesgo moral) y no existe una opción externa para las firmas (*i. e.*, no posee restricción de liquidez ni de capacidad administrativa).

Finalmente, se asume que el Estado no valora en absoluto la renta que pudiese percibir el agente y por eso lo único que considera es la renta que puede obtener del recurso. Por ello, el problema del Estado es maximizar la renta que recibe, sujeto a la restricción de participación de la firma:

$$\max \int [U_\tau - V_x(U_\tau)] dU_\tau$$

$$s.a. \int [V_x(U_\tau) - I_{t=0}] dU_\tau \geq 0$$

$$0 \leq V_x(U_\tau) \leq U_\tau$$

---

[1] El supuesto de que la inversión es igual entre todas las firmas, aunque iluso, es irrelevante: Los resultados cambiarían poco si se levanta. Contrariamente, el supuesto de no colusión es clave. La colusión de las firmas participantes puede cambiar mucho los resultados que serán expuestos (Klemperer, 2004).

La única variable de decisión del planificador es el esquema de pagos de la firma $V_x(U_\tau)$. Bajo este planteamiento, resulta claro que el esquema $V_x(U_\tau)$ que maximiza el resultado para el Estado es único e igual a la inversión, a saber $V_x(U_\tau) = I_{t=0}$ (Engel *et al.*, 2008). La implementación de este esquema de pagos puede lograrse mediante la concesión denominada Mínimo Valor Presente de Ingresos (MVPI). Una descripción detallada de ésta será presentada al final del trabajo.

En este trabajo se consideró, además de la inversión requerida para comenzar a operar el recurso, la exploración requerida para encontrarlo. Al respecto, cabe destacar que ésta no es sólo la exploración que fue exitosa, sino también aquellas que fracasaron en el intento. Esta modificación resulta pertinente para cualquier recurso que debe ser encontrado (minerales e hidrocarburos), mas no lo es para aquellos cuya existencia es conocida (peces, agua, bosques).

Así, $I_{t=0} = I_{et} + I_{er}$, en que $I_{et}$ representa la inversión para comenzar la explotación e $I_{er}$ la inversión en exploración. Ahora bien, mientras $I_{et}$ es simplemente $I_{et}$, en el caso de $I_{er}$ la situación es distinta. $I_{er} = g_{er} * p$, en que $g_{er}$ contempla el gasto de exploración incurrido para encontrar el yacimiento objeto de la concesión, y $p$, el inverso de la probabilidad de encontrar un yacimiento. Por ello, la inversión en exploración considerada para la concesión supera por varios factores el gasto efectivamente incurrido para encontrar el yacimiento específico. Esto es lo que permite separar de la renta como tal la cuasi renta requerida para incentivar la exploración de un recurso.

2. *Representación gráfica*

Este modelo puede representarse gráficamente para ilustrar la dinámica de pago de la inversión inicial. Ésta corresponde al esquema de pagos en valor presen-

Gráfica 3. *Dinámica de pago de la inversión inicial en concesiones subóptimas y óptimas*

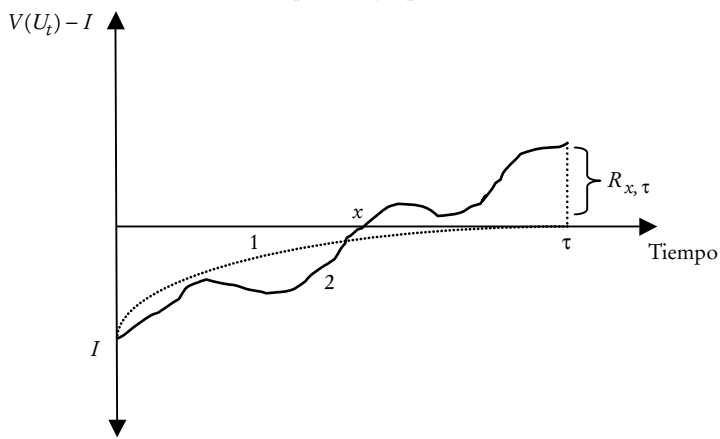

Fuente: elaboración propia.

te, acumulada en cada periodo $t$ menos la inversión inicial —$V_x(U_\tau) - I_{t=0}$—, y se denomina *renta en valor presente* (RVP). Como ya fue planteado, una concesión óptima implica que la RVP no exceda cero. Así, puede entenderse según las formas presentadas en la gráfica siguiente:

Para la concesión de un recurso que no permite la apropiación privada de renta (flujo 1), la RVP puede o no ser parsimoniosa como en la gráfica 3, siendo sólo esencial que converja en cero en algún momento lejano en que $t$ converge en $\tau$. Por su parte, en una concesión de un recurso que permite la apropiación privada de renta (flujo 2), la RVP paga la inversión inicial en un tiempo acotado, momento "x", luego del cual toda la utilidad contable que obtiene corresponde a renta apropiada por el agente.

## IV. Análisis empírico: Construcción de la RVP

### 1. *Base de datos*

El análisis empírico fue realizado con los Estados Financieros Auditados de los años 2001-2011 para cada operación relevante de un privado. Dichos estados fueron obtenidos de las Memorias anuales publicadas por la SVS (2012) y en compilados elaborados por el Consejo Minero. Las operaciones consideradas relevantes fueron aquellas que extrajeron persistentemente sobre 50 000 TMCF al año. Los datos correspondientes a gastos de explo-ración y a los años desde el inicio de cada faena hasta el año 2001 tuvieron que ser recreados.² El detalle de la metodología para recrear cada partida se detalla en la primera sección del apéndice.

Como resultado, se obtuvieron 12 bases distintas para 14 yacimientos que abarcan desde un periodo de 11 años para la faena más nueva (El Tesoro) hasta un periodo de 21 años para las faenas más antiguas (Escondida, Anglo American Norte y Sur). La muestra representó en promedio 85% de la producción privada en Chile durante el periodo 1990-2011. Los yacimientos analizados, junto con sus principales dueños, su año de apertura y la extracción promedio anual desde su apertura hasta 2011, están disponibles a solicitud, mientras que una descripción detallada de las tres variables necesarias para calibrar el modelo, se presenta en la segunda sección del apéndice.

## 2. *Evidencia Empírica*

a) *Representación gráfica.* La gráfica 4 resume toda la información relevante con una tasa de descuento de 12.36%:

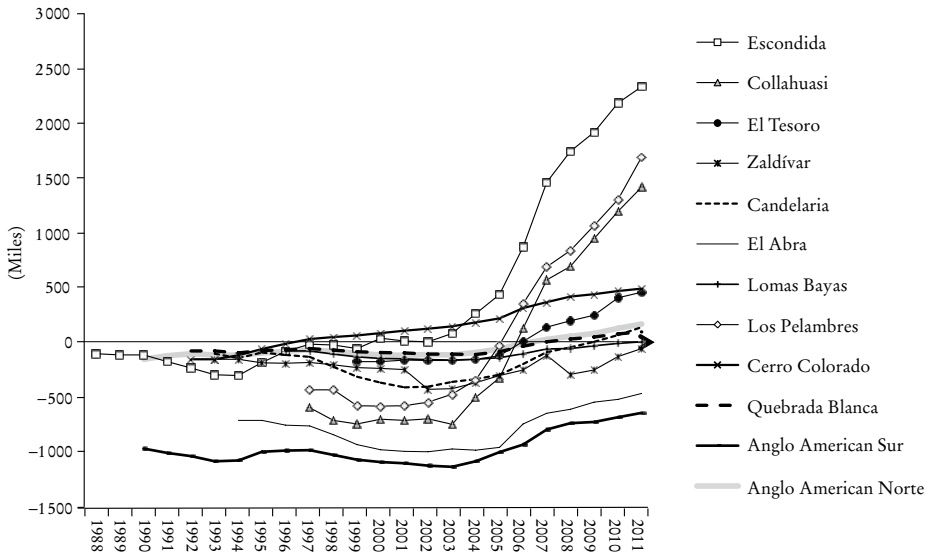

GRÁFICA 4. *Renta en valor presente*

FUENTE: elaboración propia.

---

² Los datos previos a 2001 fueron imposibles de conseguir.

Varias empresas cupríferas han percibido renta. Entre ellas destacan especialmente Escondida y Los Pelambres, seguidas por Collahuasi. Contrariamente, se puede notar que Zaldívar y especialmente El Abra y Anglo American Sur no lo han hecho. Por su parte, Lomas Bayas, Anglo American Norte y Candelaria aparecen apropiándose de una renta sutil, posiblemente sensible a la tasa de descuento.

El momento $x$ (aquel donde se comienza a percibir renta) ocurre de manera significativa para la mayoría de los yacimientos en 2005, año en que el último ciclo alto del precio del cobre se intensificó. Cabe destacar que en aquellos años la RVP de los yacimientos más rentables, como Escondida, Collahuasi y Los Pelambres, tiene un punto de inflexión. Lo anterior es esperable, si se considera que entre 2006 y 2007 la rentabilidad financiera (ROE, por sus siglas en inglés) anual promedio del GMP-10 superó 100% (Titelman, 2008).

b) *Análisis de sensibilidad.* ¿Cambiarían significativamente los resultados ante una tasa de descuento más exigente? A continuación, se presenta la RVP con una tasa de 18.78% (642 puntos porcentuales mayores):

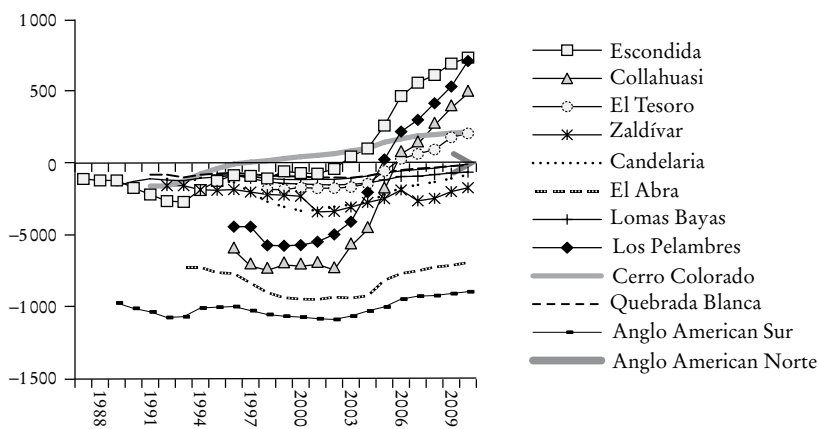

Gráfica 5. *Renta en valor presente. Caso conservador*

Fuente: elaboración propia.

Por una parte, este escenario evidencia algunos cambios, pues Lomas Bayas, Anglo American Norte, Candelaria y Quebrada Blanca ya no muestran haber percibido renta. Por otra parte, la magnitud de la apropiación de renta que muestran Escondida, Collahuasi, Los Pelambres, Cerro Colorado y El Tesoro sigue siendo considerable en este escenario conservador.

Otro resultado interesante es que el momento $x$ se retrasa para algunas mineras, mientras que para otras no. Aquellas sensibles a la tasa de descuento son Escondida (2000 a 2004); Collahuasi (2006 a 2007); Cerro Colorado (1997 a 1998) y, con mayor notoriedad, Anglo American Norte (2007); Quebrada Blanca (2007); Candelaria (2009), y Lomas Bayas (2011), pues dejan de mostrar apropiación de renta. Al otro extremo, llama la atención que el momento $x$ no cambie para El Tesoro (2006) y Los Pelambres (2006).

Probablemente lo más llamativo sea la robustez frente a la flexibilización de la tasa de descuento: la considerable renta que muestra haberse apropiado Escondida, Los Pelambres y Collahuasi y la ocurrencia del momento $x$ en múltiples yacimientos durante el último ciclo alto del precio del cobre.

c) *El modelo.* El modelo presentado anteriormente permite cuantificar la renta que ha sido apropiada por las empresas privadas. Para cada agente se conoce tanto la magnitud óptima del valor presente del esquema de pagos $V_x$ que debiese haber recibido —el valor de la inversión inicial—, así como la utilidad contable de cada proyecto $U_{2011}$ hasta el 2011, y por tanto se puede estimar la renta en valor presente $U_{2011} - V_x$ que ha percibido cada empresa. Adicionalmente, también se puede aproximar el valor corriente de dicha renta.

En el cuadro 1 se presenta la magnitud de la renta en valor presente para cada yacimiento en $t = 0$, así como el monto que correspondería al año 2012 asumiendo que ésta hubiese sido ahorrada en el Fondo de Estabilización Económica y Social (FEES).[3] Este ejercicio permite cuantificar lo que ha dejado de percibir el Estado chileno por poseer una institucionalidad minera insuficiente.

Cuadro 1

| Yacimiento | Momento x | | Renta apropiada en VP (millones de dólares) | | Renta apropiada a 2012 (millones de dólares de 2012) | |
|---|---|---|---|---|---|---|
| | 12.36% | 18.78% | 12.36% | 18.78% | 12.36% | 18.78% |
| Escondida | 2000 | 2004 | 2 329.8 | 734.7 | 27 628 | 25 624 |
| Los Pelambres | 2006 | 2006 | 1 688.4 | 711 | 6 379 | 6 379 |
| Collahuasi | 2006 | 2007 | 1 414.4 | 486.3 | 5 980 | 4 170 |
| Candelaria | 2009 | — | 136 | — | 1 094 | — |
| A. A. N. | 2007 | — | 166.3 | — | 1 561 | — |
| C. Colorado | 1997 | 1998 | 489 | 218.2 | 3 119 | 2 874 |
| El Tesoro | 2006 | 2006 | 457.8 | 208 | 1 673 | 1 673 |
| Q. Blanca | 2006 | — | 92.8 | — | 771 | — |
| Lomas Bayas | 2011 | — | 4.9 | — | 0[a] | — |
| Total | | | | | 48 208 | 40 722 |

[a] Elaboración propia. Como se consideraron flujos después del año de pago de la inversión inicial, no se pudo valorar la renta que ha generado

[3] La rentabilidad neta en dólares anualizada de dicho fondo ha sido 5.07% anual desde el 31 de marzo de 2007 (Hacienda, 2011). Se consideraron los flujos usados para construir la RVP, desde el año siguiente al momento $x$ y se multiplicaron por dicho interés compuesto hasta 2012.

El cuadro 1 permite apreciar detalladamente los resultados ya expuestos en el apartado anterior. El costo —en valor de 2012— de haber cedido la renta de Escondida ascendió a 27.6 o 25.6 mil millones de dólares, seguido por la de Los Pelambres con 6 300 y Collahuasi con 6 000 o 4 000 millones de dólares. La tabla también muestra que de las 12 empresas analizadas, nueve ya han pagado en valor presente su inversión inicial con una tasa de descuento de 12.36%, y cinco lo han hecho con una de 18.78%.

## V. Resultados y políticas

### 1. *Implicaciones de los resultados*

En términos proporcionales, el costo agregado entre 48.2 y 40.7 mil millones de dólares por haber cedido la renta del cobre a las principales empresas cupríferas durante los últimos 10 años —a lo cual Escondida aporta entre 57 y 63%— representa entre 91 y 77% de los Ingresos Generales de la Nación considerados para la elaboración del presupuesto 2012[4] (DIPRES, 2012), o alternativamente entre 2.7 y 2.3 veces el valor de mercado conjunto del FEES y el FRP al año 2011 (Hacienda, 2011). Sin embargo, lo más importante de los resultados no es la magnitud sino su robustez. Las conclusiones cambian sólo levemente ante un incremento de la tasa de descuento de 642 puntos porcentuales.

Una potencial limitación del análisis es que está acotado al año 2011, lo que tiene dos implicaciones. La primera es que no reconoce el conjunto de pasivos de corto y largo plazo que cada empresa posee y deberá pagar en el futuro, mientras que la segunda es que no se hace cargo del conjunto de la utilidad contable que seguirá percibiendo cada empresa durante la vida útil de los yacimientos.

A menos que se estime que el precio del cobre se desplomará a tal nivel que lleve a la insolvencia a las mineras, lo anterior no afecta las conclusiones obtenidas. La única forma en que estas conclusiones cambiarían es si las mineras comienzan a operar a pérdida y deciden no cerrar. Resulta dudoso que Escondida esté dispuesta a perder 27 000 millones de dólares antes de simplemente terminar la faena.

Los 40.89 mil millones de dólares que corresponderían a la renta, que el Estado de Chile ha dejado de percibir debido a la inadecuada institucona-lidad vigente, indicarían, por lo tanto, que ha existido apropiación privada de renta en este país. En caso de ser correcto este análisis, la renta que han percibido las principales mineras privadas en Chile es evidencia suficiente de que la falta de sofisticación de la institucionalidad minera ha provocado dos graves problemas, y sugiere la posibilidad de un tercero.

---

[4] En la transformación de cifras se consideró una tasa de cambio de 500 pesos por dólar.

Al ceder la renta del cobre, el Estado perjudica la eficiencia del sistema económico. Por una parte desperdicia una benigna y considerable fuente de recaudación fiscal, con la cual se podrían disminuir otros impuestos distorsionadores, entre otras cosas. Por otra, permite que mineras ineficientes puedan sobrevivir en el mercado por un tiempo considerable: el principal mecanismo para discriminar entre firmas eficientes e ineficientes ya no es válido y, por tanto, no se puede saber si Escondida, Los Pelambres y Collahuasi sobreviven en el mercado por una gestión eficiente o tan sólo por la renta que gozan.

Al no captar la renta del cobre, el Estado también ocasiona problemas de equidad al otorgar retornos extraordinarios de manera arbitraria a unos pocos individuos. Un análisis adecuado de esto escapa del alcance de este trabajo, pero resulta sugerente que la familia más rica del país sea dueña de Los Pelambres y que varias de las que le siguen estén relacionadas a otros recursos como peces y bosques.

Por último, ceder la renta del cobre, en particular a empresas extranjeras que repatrian una proporción considerable de sus utilidades (Escondida pagó dividendos por 10 700 millones de dólares entre 2006 y 2007), pone una interrogante a la perspectiva de poder reemplazar al cobre una vez que se agote.

Cabe aclarar dos puntos respecto a las implicaciones de los resultados. Uno es que la evidencia de apropiación privada de renta no induce a una crítica contra las concesiones de recursos *per se*; el Estado de Chile difícilmente podría haber financiado la magnitud de inversiones requeridas durante la década de 1990 en este sector, y en dicho contexto las concesiones son un mecanismo idóneo. Por ello, el problema de apropiación de renta radica en el diseño de la concesión y no en el uso de una concesión en sí. El segundo punto es que estos resultados se encuentran naturalmente sujetos a los supuestos utilizados, entre los cuales destacan dos: el proyecto siempre es rentable y no existe riesgo de expropiación. Respecto al primero, lo que la teoría indica es que la concesión debiese incorporar un ingreso mínimo garantizado para aquellos casos en que el proyecto resulte no rentable (En-gel *et al.*, 2013). No obstante, en este trabajo abordamos esta situación con la tasa de descuento. Haber incluido para el caso conservador un coeficiente Beta de dos en la tasa de descuento tiene el sentido de considerar que existe un riesgo real para la actividad, pues puede no ser rentable. Algo similar sucede con el riesgo de expropiación: al considerar un riesgo soberano de 411 puntos, más de dos veces superior al riesgo soberano que mostraba Chile en 1999 cuando comenzaba su serie en el Indicador de Bonos de Mercados Emergentes y cercano al valor de Argentina, se está insinuando en el modelo que invertir en el país supone un riesgo proveniente de la posibilidad de expropiación, entre otras cosas. Si para todos los Estados del mundo posibles la actividad fuese rentable y no existiese riesgo de expropiación, la tasa de descuento no debería incorporar riesgo alguno y, por tanto, tomaría el valor del retorno libre de riesgo. Resulta difícil llegar a un valor certero para ambas variables, y es posible plantear que debiesen ser mayores, en cuyo caso la cifra de renta disminuiría. No obstante los valores presentados son conservadores.

Resulta interesante notar que este análisis puede utilizarse para evaluar la apropiación de renta en cualquier país que cuente con suficiente información acerca de sus faenas de explotación de recursos naturales, lo que resulta prometedor para varias naciones latinoamericanas altamente orientadas a la extracción de minerales, hidrocarburos, peces u otros recursos naturales. El presente análisis puede ser extendido a cualquier recurso natural que cuente con las características básicas de requerir una inversión inicial definible, tener una utilidad contable y rastreable y poseer una tasa de descuento estimable.

## 2. Recomendaciones de política

Dada la sugerente evidencia de que diversas empresas privadas se han apropiado de renta del cobre, las propuestas de realizar cambios en la institucionalidad vigente resultan persuasivas: el análisis realizado arroja luces de las características que debe tener. Esta experiencia puede ser útil para el diseño de concesiones en cualquier país.

Si bien existe una extensa lista de instrumentos disponibles para captar renta (Otto, 1992 y 2000; Boadway y Flatters, 1993; Eggert, 1999; Baunsgaard, 2001; Dietsche *et al.,* 2009; Freebairn y Quiggin, 2011; Guj, 2012), sólo nos concentraremos en el diseño de concesiones. La apropiación privada de la renta de recursos ocurre en muchos casos porque las concesiones no están diseñadas con la consideración de que puedan obtenerse ganancias extraordinarias. De hecho, en Chile, la Ley Orgánica de Concesiones Mineras fue explícitamente diseñada asumiendo que no existe renta (Piñera, 2002).

Una concesión óptima de recursos naturales, relativa al manejo de renta, debe partir por ser lo suficientemente flexible para ajustarse a los ciclos de precios propios de estos sectores y para compensar desviaciones de uno de sus parámetros con ajustes de otro. Debe requerir la mínima cantidad de información por parte del Estado y del agente y depender de variables fácilmente observables e idealmente exógenas a los involucrados; debe procurar competencia *por la cancha* y evitar incentivos perversos en el comportamiento del agente una vez asignada la concesión.

Al respecto, el MVPI tiene muchas características atractivas (Engel *et al.*, 1998 y 2008). El Estado licita un recurso al indicar sus características relevantes —localización, abundancia, etc.— e indicar la tasa de descuento con la cual se descontará la utilidad contable futura. La concesión se asigna al participante que exija el menor valor presente de ingresos a cambio de realizar la inversión inicial necesaria para extraer el recurso. El ganador de la licitación se asigna el derecho exclusivo de explotación hasta el momento en que obtiene el VPI ofrecido. En ese momento, la concesión expira. Así, esta concesión es una operacionalización del marco teórico usado en este trabajo.

El MVPI tiene la característica de que, si la tasa de descuento definida por el Estado es baja (relativa al riesgo de que el proyecto no sea rentable o que la inversión sea expropiada), los participantes simplemente ajusten hacia arriba su VPI ofrecido conforme al cumplimiento con su restricción de participación.

Un punto que sobresale es que esta forma de concesión implica que el proceso de exploración ya debió haberse realizado y se requieran mecanismos específicos para incentivar dicha actividad. Uno de ellos podría ser, al estilo mexicano, el establecimiento de un servicio público encargado de explorar; otro podría ser la vinculación de las exploraciones privadas exitosas a los flujos futuros de los yacimientos encontrados. El mecanismo más conveniente es un tema de futura investigación.

Toda vez que se asume que los participantes no se coluden, que pueden prever adecuadamente sus costos futuros y que los riesgos políticos como cambios en la política tributaria o expropiación son incorporables en su tasa de descuento, entonces por equilibrio de Nash éstos ofrecerán el MVPI que logre pagar la inversión inicial. La lógica detrás del monto ofertado es que, sustraídos todos los costos, la explotación deja una utilidad contable que en valor presente se iguala a la inversión inicial.

Las propiedades de esta forma de concesión son múltiples:

*i)* El MVPI es de plazo contingente al precio. Esto elimina para el agente el riesgo de demanda, lo que resulta óptimo para una concesión al ser una variable que el operador no puede controlar (Irwin, 2007). Así, el plazo de la concesión se ajusta al precio del recurso, con lo cual asegura un retorno adecuado al capital y redirige todo exceso al Estado.

*ii)* El MVPI tiene bajos requerimientos de información: el Estado puede ignorar la magnitud de la inversión inicial así como el costo operacional. Tanto el Estado como el concesionario pueden ignorar la función densidad del precio. El hecho de que el plazo se ajuste según el precio y se compense con la tasa de descuento torna dicha información irrelevante.

*iii)* La concesión depende del ingreso, variable fácilmente observable: está definida por el precio, usualmente fijado en grandes mercados (en el caso del cobre, en la Bolsa de Metales de Londres —BML—) y por la cantidad, variable típicamente fiscalizada en aduanas, puertos o agencias especializadas (nuevamente en el caso del cobre, por Aduanas de Chile y la Corporación Chilena del Cobre).

*iv)* Se trata de una licitación que discrimina entre empresas eficientes e ineficientes, pues naturalmente una empresa ineficiente tendrá que ofrecer un VPI mayor para recuperar la inversión.

*v)* La licitación neutraliza mecanismos elusivos como precios de transferencia. Como dichos mecanismos operan abultando los costos de las empresas, la concesión MVPI fuerza por equilibrio de Nash a que los participantes consideren sus mínimos costos posibles de operación.

*vi)* La concesión puede tener una cláusula que establezca un impuesto especial —voluntario— con el cual se reduzca el ingreso contabilizado en el periodo en relación uno a uno. Esto amortigua el crecimiento del VPI y alarga la concesión.[5]

*vii)* La concesión permite valorar el monto de indemnización en caso de expropiación. Éste correspondería —en valor presente— a la diferencia entre el VPI ofertado y el ingreso en valor presente percibido hasta el momento de la expropiación. Aun así, la posibilidad de cuantificar la indemnización no asegura que ésta sea efectivamente pagada por el Estado, y así no elimina este riesgo con variados precedentes en países latinoamericanos. Una solución a ello es la opción tomada por Chile, en la que el Estado da garantías mediante la suscripción de un Contrato Ley con el agente. Otra es que la inexistencia de credibilidad intertemporal de los compromisos adquiridos simplemente se traduzca en una mayor tasa de descuento.

---

[5] Esta posibilidad de la concesión MVPI resulta evidente, aunque no ha sido presentada en los trabajos de Engel *et al.* (2001 y 2008).

APÉNDICE

1. *Estimación de valores previos a 2011*

a) *Ingresos de explotación:* precio del cobre (BML) de cada año, por la cantidad producida o exportada en cada año, según fuese aquel que diera un resultado menor.

b) *Costos de explotación:* promedio del costo por tonelada producida de los últimos cinco años (2001-2005), multiplicado por la cantidad producida en cada periodo.

c) *Gastos de administración y ventas:* promedio de la razón entre el gasto de administración y ventas y el costo de explotación de los últimos cinco años (2001-2005), multiplicando dicha razón por el costo de explotación de cada periodo.

d) *Resultado operacional:* diferencia entre el margen operacional y los gastos de administración y ventas de cada periodo.

e) *Resultado no operacional, incorporación de activos fijos, depreciación y amortización y pago de préstamos neto:* se consideró el promedio de los últimos cinco años (2001-2005).

f) *Pago de impuestos:* durante el periodo 1990-2000 muy pocas mineras pagaron impuestos. Sólo se consideraron los pagos de impuestos de Escondida, mientras que para todas las otras mineras éstos se asumieron iguales a cero.

g) *Imputación de gastos de exploración:* primero, los gastos de exploración total de Chile fueron transformados de un valor porcentual sobre PIB (formato de la serie original del Banco Central de Chile) a valores en dólares corrientes con series de PIB del Banco Mundial.

Hay que mencionar que: *i)* los gastos de exploración privado y público fueron considerados según la proporción de producción de largo plazo *(i. e.,* del año 2011); por lo tanto, dos tercios y un tercio de los gastos fueron atribuidos al sector privado y público respectivamente; *ii)* los gastos privados en exploración en un año dado fueron prorrateados entre toda minera que estuviese en un periodo de cinco a cero años de realizar su inversión inicial. El prorrateo se realizó en proporción a la producción media de cada minera, sobre la producción total media de todas las mineras consideradas para dicho año, y *iii)* el valor resultante fue actualizado para considerar su costo de oportunidad. La tasa de descuento usada fue la misma utilizada en cada escenario.

2. *Variables clave*

a) *Inversión inicial:* corresponde a la partida "capital pagado" del primer año del que se tuviese información (generalmente 2001 con excepción de Escondida que se tuvo para el año 2000), y se asumió que éste fue el monto original aportado por los inversionistas. Esto es una aproximación conservadora, pues todas las mineras exhiben una tendencia creciente de dicha partida con los años.

Una aproximación adicional fue usada para incluir los gastos en exploración. Se utilizó la metodología descrita antes para imputar los ocho yacimientos que comenzaron su operación durante el periodo en análisis la totalidad del gasto privado en exploración minera durante 1984-1999. Dicha aproximación fue utilizada por la disponibilidad de datos y por su equivalencia con el modelo presentado. Tal equivalencia deriva de:

$$I_{er} = g_{er} * p \to I_{er} = g_{er} * \frac{1}{\frac{n}{T}} \to I_{er} = g_{er} * \frac{T}{n} \to I_{er} = g_{er} * t \to I_{er} = g_{ert}$$

En que $n$ representa el número de campañas de exploración exitosas; $T$, el número total de campañas de exploración realizadas; $t = \tau/n$ la cantidad de campañas realizadas por cada campaña exitosa; y $g_{ert}$, el gasto total en exploración realizado. Por ello, esta equivalencia se sostiene sólo en tanto que el gasto del yacimiento específico $g_{er}$ sea igual al gasto en exploración promedio. Así, los yacimientos bajo explotación cubren la inversión requerida para poder extraer minerales, los gastos de exploración para encontrar el yacimiento y los gastos de exploración de todas las campañas de exploración sin éxito para mantener los incentivos a continuar con futuras exploraciones.

b) *Utilidad contable:* corresponde al flujo de caja para proyectos mineros, incluyendo la partida "pago de préstamos neto". La lógica de esto es considerar el flujo neto de caja del proyecto en cada periodo.

Por último, se asume que toda la información aportada por las mineras en sus estados Financieros es precisa y que no hay distorsiones por precio de transferencia o sobre endeudamiento con partes relacionadas. Las fuentes de los datos utilizados están disponibles ante requerimiento.

El siguiente cuadro resume los componentes del flujo:

Cuadro A1

|  | *Utilidad contable* |
|---|---|
| Se suman | Resultado antes de impuestos |
|  | Depreciación y amortización |
| Se restan | Aumento de capital oagado |
|  | Pago de impuestos |
|  | Incorporación de activos fijos |
|  | Pago de préstamos neto |

Fuente: elaboración propia del autor.

Sobre impuestos, se incluyen el Impuesto de Primera Categoría, Específico a la Actividad Minera (desde 2006), y Adicional.

c) *Tasa de descuento:* se consideró un Modelo de Valoración de Activos Financieros (capm, por sus siglas en inglés) (Sharpe, 1964) más riesgo soberano. Ésta no es la única metodología para estimar el costo de capital, pero es la dominante (Bruner *et al.*, 1998). No se utilizó un modelo de Costo del Capital Medio Ponderado (wacc, por sus siglas en inglés), pues sólo se está considerando como inversión el capital efectivamente aportado por los socios. Por tanto:

$$c = \text{tasa libre de riesgo} + \beta (\text{premio por patrimonio}) + \text{riesgo país}$$

La tasa libre de riesgo utilizada fue de 6.9%; el Beta, de 0.91; el premio por patrimonio, de 3 889%; y el riesgo país, de 1.73%. La imputación de estos datos en un modelo capm da una tasa de 12.36%.

Por su parte, la utilización de un $\beta$ de 2 y la consideración de una prima por riesgo soberano de 411 puntos básicos da una tasa de descuento de 18.78%.[6] Un $\beta$ de 2 es superior a las magnitudes encontradas en Reuters, Damodaran, WikiWealth, Bloomberg, y Stock Analysis on Net para la industria minera. Un riesgo soberano de 411 puntos básicos significa otorgarle el máximo riesgo de la serie del embi, correspondiente al 23 de enero de 2009.

Por último, la tasa de descuento usada es nominal y, en consecuencia, tiene el efecto de inflación incorporado. Esto supone que no es necesario aplicar posteriormente algún tipo de deflactor a las cifras (Sapag, 2008).

Referencias bibliográficas

Banco Mundial (2010), Wealth of Nations Dataset, disponible en http://data.worldbank.org/data-catalog/wealth-of-nations

---

[6] Esto está en línea con otras estimaciones del costo de patrimonio de la industria minera. Como ejemplos, WikiWealth publica el capm de Rio Tinto (13.4), bhp Billiton (12.2), Vale (12.9) y Anglo American (13). Por su parte el Banco ubs considera una rentabilidad de 15% adecuada para el sector.